\documentclass{iau}
\usepackage{graphicx}

\title[X-ray superclusters] 
{Characterising large-scale structure with \\ the REFLEX II cluster survey}

\author[Gayoung Chon]   
{Gayoung Chon$^1$
}

\affiliation{$^1$Max-Planck-Institut f{\"u}r extraterrestrische Physik \\ Garching 85748,
Germany \\ email: {\tt gchon@mpe.mpg.de} \\[\affilskip]
}

\pubyear{2014}
\volume{308}  
\pagerange{xxx--xxx}
\jname{THE ZELDOVICH UNIVERSE}
\editors{R. van de Weygaert, S. Shandarin, E. Saar \& J. Einasto, eds.}
\begin{document}

\maketitle

\begin{abstract}
We study the large-scale structure with superclusters from the REFLEX X-ray cluster
survey together with cosmological N-body simulations.
It is important to construct superclusters with criteria such that they are homogeneous 
in their properties. 
We lay out our theoretical concept considering future evolution of superclusters
in their definition, and show that the X-ray luminosity and halo mass functions of clusters
in superclusters are found to be top-heavy, different from those of clusters in the field. 
We also show a promising aspect of using superclusters to study the local cluster bias
and mass scaling relation with simulations.
\end{abstract}
\firstsection 
\section{Introduction}

Superclusters are the largest overdense structures found in the galaxy and galaxy 
cluster surveys, defined in general as groups of two or more galaxy clusters above 
a certain density enhancement (\cite{bahcall88}). 
Unlike clusters they have not reached a quasi-equilibrium configuration, hence 
the definition of superclusters must be made clear such that their properties
can be studied quantitatively. 
One solution to this problem is to include the future evolution of superclusters 
into the definition of the object, selecting only those structures that will 
collapse in the future.
This allows us to obtain a more homogeneous class of superclusters. 
We have been exploring this approach observationally using an appropriate selection criterion 
to construct an X-ray supercluster catalogue (\cite{gc13,gc14}). 
It is based on the REFLEX X-ray cluster survey, which provides the largest and homogeneous 
X-ray cluster sample to date in the southern sky (\cite{b13,gc12}). 
Since the REFLEX survey has a well-defined selection function, we can apply equivalent 
criteria to dark matter halos in cosmological N-body simulations to construct superclusters, 
which allows us to study superclusters more quantitatively. 

\section{Defining superclusters}

To identify a region that will collapse in the future, we approximate the
overdense regions by homogeneous density spheres, which has been 
successfully used in the literature for many applications.
We can then model the evolution of the overdense region with respect to 
the expansion of the background cosmology with reference to Birkhoff's theorem.
This allows us to describe the evolution of both the overdense and background
regions by the respective values of the local and global Hubble constant,
H, matter density, $\Omega_m$, and $\Omega_{\rm \Lambda}$ corresponding to 
a cosmological constant.
We evolve both regions from a starting redshift of 500, which results in an 
accuracy well below 1\% in the final calculation.
We solve the Friedmann equations for a spherical collapse model iteratively where 
the local matter density is enhanced at the starting redshift in the overdense region. 
We then obtain an criterion for R, which is the ratio between the minimally required
local overdensity to the mean density of the universe today. 
This density ratio, for example, is 7.858 for the flat $\Lambda$CDM cosmology with 
$\Omega_m$=0.3 and $h$=0.7.
Since we use a friends-of-friends (fof) algorithm to build superclusters from clusters, 
the linking length must reflect this required density ratio. 
Given that the linking length is inversely proportional to a third power of the local density 
of clusters, we find that the overdensity parameter, $f$, used in the fof algorithm 
has to be about 25 for a cluster bias of 2-3.
In~\cite{gc13} we adopted a slightly more generous value of ten for the nominal 
catalogue together with that built with $f$=50 for comparison so as to collect a slightly 
larger regions than just the core of the collapsing superclusters. 
We recover a number of known superclusters including Shapley, Hydra-Centaurus, and Aquaris B
as well as a number of new X-ray superclusters. 
Based on our physically driven choice of the overdensity parameter, one of the consequences
is that, for example, the Shapley supercluster is fragmented into three smaller mass concentrations.
A further study of the radial profile of Shapley mass concentration implies that 
in fact approximately the central 11~h$^{-1}$\,Mpc of Shapley is under-going a collapse currently,
while regions outside 13~h$^{-1}$\,Mpc will not collapse in the future despite the fact 
that the outskirts of Shapley are rich with clusters (Chon et al., submitted).

\section{Superclusters as dark matter tracers}

The fact that the REFLEX II supercluster sample has been constructed by means of
a statistically well-defined sample of closely mass-selected clusters motivates
us to search for a more precise physical characterisation of the superclusters
in simulations. 
We achieve this by applying criteria equivalent to those used in our X-ray selection,
and regard halos as clusters by imposing a mass limit to halos in the Millennium 
simulation (\cite{springel}). 
Our nominal lower mass limit is 10$^{14}$h$^{-1}$M$_{\odot}$ corresponding to a typical
mass limit of a cluster survey, and we construct superclusters with the same
overdensity parameter that was used for the REFLEX II superclusters. 

One of the interesting aspects in the study of superclusters is how superclusters trace
the underlying dark matter. 
Since we have almost no direct access to determined their masses, an indirect mass
estimate would still be helpful.
Since we have mass estimates of the member clusters through mass-observable scaling 
relations, the total supercluster mass can be estimated if we can calibrate
the cluster mass fraction in superclusters.
Similarly we would be able to determine the dark matter overdensity traced by
a supercluster if the mass fraction relation or the overdensity bias could be
calibrated.
Hence we consider two quantities, the mass fraction represented by the total clusters 
mass in a supercluster, and the bias of clusters in superclusters. 
A better knowledge of the former would allow us to make a prediction of the 
underlying total mass of a superclusters through a mass scaling relation
like that of clusters. 
\begin{figure}[bt]
  \begin{center}
    {\includegraphics[width=0.49\columnwidth]{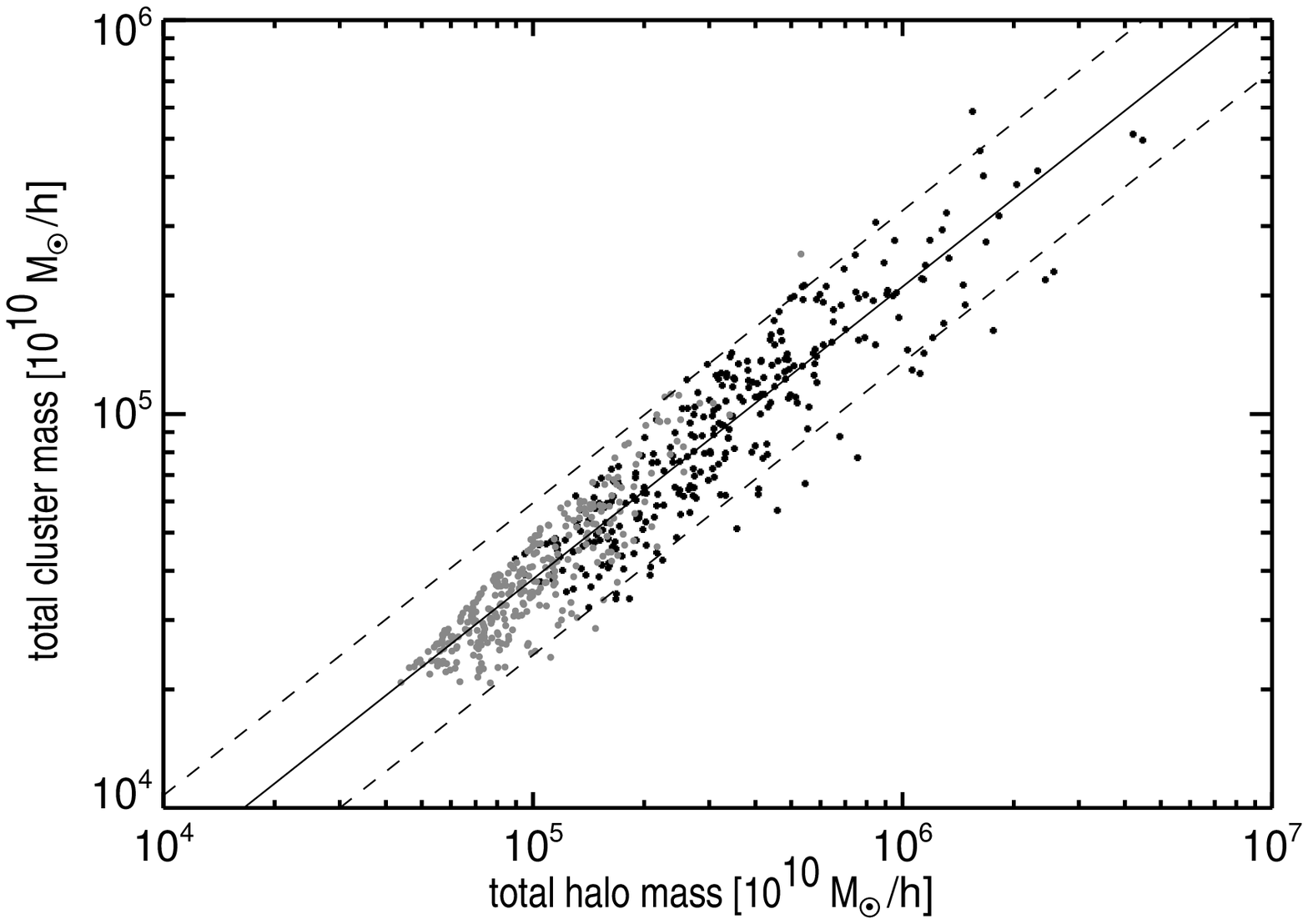}}
    {\includegraphics[width=0.49\columnwidth]{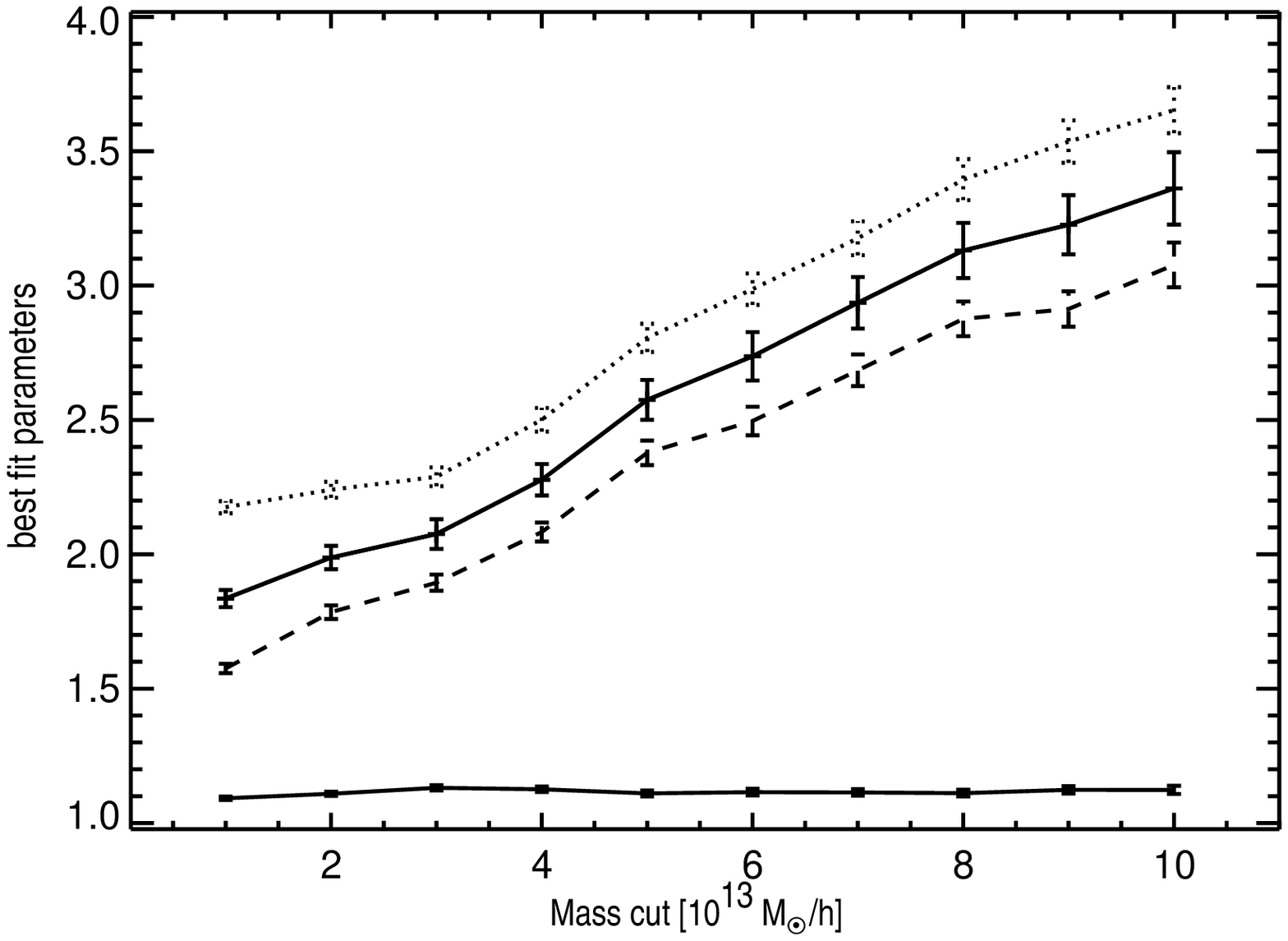}}
    \caption{
      (Left) Total mass of a supercluster probed by the total mass of member clusters. 
      The halos of the lower halo mass limit of 10$^{14}$h$^{-1}$M$_{\odot}$ 
      are considered here.
      (Right) Fitted slope and amplitude for the mass bias as a function
      of the mass limit in the cluster catalogue. In comparison we show 
      the fitted amplitude of the pair (dotted) and richer (dashed) superclusters
      separately where the slope is fixed to that for the entire sample.} 
    \label{fig:mfrac-bias}
  \end{center}
\end{figure}

The left panel of Fig.~\ref{fig:mfrac-bias} shows the total cluster mass as a 
function of the total mass of superclusters with the best fit to a power-law shown 
in a solid line, which is enclosed by two dashed lines representing its two sigma scatter.
The total mass of a supercluster is defined to be the sum of all halos in the volume
with a correction factor that takes the particles which do not contribute to the 
halo mass, M$_{\rm 200}$, assuming that they are distributed throughout the volume in an unbiased way.
The fact that the total cluster mass is closely correlated with the total halo
mass of a supercluster gives a first encouraging evidence that there is a potential
to use the cluster observables to trace the dark matter distribution in the regions
of superclusters.

The power spectrum of clusters of galaxies measures the density fluctuation amplitude of 
the distribution of clusters as a function of a scale, where the amplitude ratio of this
power spectrum in comparison with the power spectrum of the dark matter is
interpreted as a bias that clusters have.
Analogous to the power spectrum of clusters, we take the number overdensity
of clusters in superclusters as a measure of bias against the dark matter 
overdensity, for which we take again the halo mass overdensity as a tracer.
This approach makes use of the observable, the number overdensity of clusters,
so it can be calibrated against a quantity from simulations, the dark matter 
mass overdensity. 
In this case we consider a continuous range of lower halo mass limits from $10^{13}$h$^{-1}$M$_{\odot}$
to $10^{14}$h$^{-1}$M$_{\odot}$ to form ten cluster catalogues from the simulation
and construct ten supercluster samples. 
We fit a power-law to the cluster number density as a function of the mass density 
of halos in superclusters. The best fit parameters are shown in the right panel of
Fig.~\ref{fig:mfrac-bias} where dotted and dashed lines represent those superclusters 
with two member clusters and richer superclusters. The uncertainties are calculated
by one thousand bootstrappings of the sample. 
The fitted slope of this relation turns out to be nearly constant, $\sim$1.1, for 
all mass ranges of consideration, hence we interpret the fitted amplitude as a bias. 
We see that the bias increases with an increasing lower mass limit in a cluster
catalogue, which is also expected. 
We note that for the $10^{14}$h$^{-1}$M$_{\odot}$ mass limit, the local bias
is 3.36, and for the $10^{13}$h$^{-1}$M$_{\odot}$ it decreases to 1.83. 
This result is in line with the biases obtained from the power spectrum analysis,
the former being 2.1 and the latter 3.3 (\cite{r2ps}).
The fair agreement between the cluster biases calculated locally and globally
is encouraging because the cluster number overdensity clearly extends into
non-linear regime whereas the calculated bias of the power spectrum is
mainly based on linear theory. This motivates the next section where we will
quantitatively test how much this non-linear environment differs from the field.

\section{Supercluster environment}

\begin{figure}[bt]
  \begin{center}
    \includegraphics[width=3.4in]{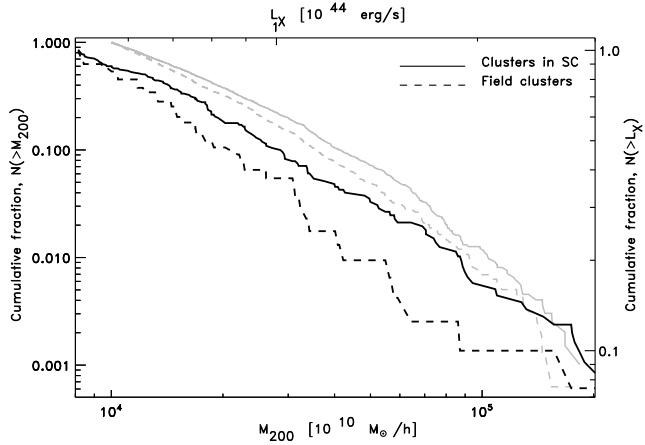} 
    \caption{Measured cumulative X-ray luminosity function from REFLEX
      shown in thick black lines in comparison to the cumulative mass 
      function of the superclusters in simulations in grey lines.
      In both cases clusters in superclusters are denoted by a solid line,
      while field clusters by dashed lines.
      }
    \label{fig:lx}
  \end{center}
\end{figure}

Cluster mass functions can be predicted on the basis of cosmic structure
formation models such as the hierarchical clustering model for CDM universe
(\cite{press74}). 
Due to the close relation of X-ray luminosity and gravitational mass of clusters
the mass function is reflected by the X-ray luminosity function (XLF) of clusters.
Thus we use the observed XLF as a substitute for the mass function.
The clusters in superclusters are used separately to form their own XLF
in comparison to those in the field, and to better illustrate the difference
we consider a normalised cumulative luminosity function as shown in Fig.~\ref{fig:lx}.
The black lines represent the volume-limited sample of the REFLEX II clusters.

We see a clear effect that we have a more top-heavy luminosity function for
the clusters in superclusters compared to those in the field. 
Since the cumulative distribution is unbinned, we quantify the difference
with a Kolmogorov-Smirnov (KS) test yielding the probability, P, 
that both luminosity distributions come from the same parent distribution
with a value of 0.03. 
With the Millennium simulation, we form the same cumulative distribution directly 
with masses shown in grey lines in Fig.~\ref{fig:lx}. 
In this case the probability from the KS test is practically zero.
Hence both in our REFLEX data and simulation that there are over-abundance
of more luminous clusters in the superclusters than in the field.



\end{document}